\documentclass{article}
\usepackage[utf8]{inputenc}
\usepackage{graphicx}
\usepackage{amsfonts, amsmath, amsthm, amssymb}
\usepackage{listings}
\usepackage{xcolor}
\usepackage[margin=1in]{geometry}

\newcommand{\bb}[1]{\mathbb{#1}}
\newcommand{\Cl}{C\ell}
\newcommand{\CCl}{\mathbb{C}\ell}
\date{}
\begin{document}

\title{Three generations of colored fermions with $S_3$ family symmetry from Cayley-Dickson sedenions}
\maketitle
\begin{center}
\author{Niels G. Gresnigt\textsuperscript{1$\star$},
Liam Gourlay\textsuperscript{1} and
Abhinav Varma\textsuperscript{1}}
\end{center}

\begin{center}
{\bf 1} Department of Physics, School of Mathematics and Physics, Xi'an Jiaotong-Liverpool University, Suzhou, China
\\
* niels.gresnigt@xjtlu.edu.cn
\end{center}

\section*{Abstract}
An algebraic representation of three generations of fermions with $SU(3)_C$ color symmetry based on the Cayley-Dickson algebra of sedenions $\mathbb{S}$ is constructed. Recent constructions based on division algebras convincingly describe a single generation of leptons and quarks with Standard Model gauge symmetries. Nonetheless, an algebraic origin for the existence of exactly three generations has proven difficult to substantiate. We motivate $\bb{S}$ as a natural algebraic candidate to describe three generations with $SU(3)_C$ gauge symmetry. We initially represent one generation of leptons and quarks in terms of two minimal left ideals of $\bb{C}\ell(6)$, generated from a subset of all left actions of the complex sedenions on themselves. Subsequently we employ the finite group $S_3$, which are automorphisms of $\bb{S}$ but not of $\bb{O}$ to generate two additional generations. Given the relative obscurity of sedenions, efforts have been made to present the material in a self-contained manner.


\section{Introduction}\label{sec:Introduction}
Despite its great practical success in colliders and other experiments, there are several unexplained features of the Standard Model of particle physics (SM) which lack a deeper theoretical motivation. These include, among others, a derivation of the SM gauge group from first principles, an explanation for why some representations of the SM gauge group correspond to particle multiplets whereas others do not, and an account for why fermions come in three generations. These theoretical shortcomings may be suggestive that the SM ultimately emerges from a more fundamental physical principle or mathematical structure.

In an attempt to establish the geometric and algebraic roots of the SM, several proposals have been put forth over the years which take as its essential mathematical ingredients (tensor products of) the only four normed division algebras over the reals: $\bb{R}$, $\bb{C}$, $\bb{H}$, and $\bb{O}$. Instead of unifying the internal symmetries into a single larger group, as is done in grand unified theories (GUTs) such as $SU(5)$ and $Spin(10)$, these division algebraic approaches attempt to unify the gauge groups together with the leptons and quarks that they act on into a single unified algebraic framework, in terms of an algebra acting on itself.

The octonions $\bb{O}$, the largest of the division algebras, were first considered in the $70$s for their intriguing efficacy in describing quark color symmetry \cite{gunaydin1973quark}. Dixon \cite{dixon1990derivation,dixon2004division,dixon2013division} considers the algebra $\bb{R}\otimes\bb{C}\otimes\bb{H}\otimes\bb{O}$ and its invariant subspaces in connection to the particles and charges of the SM. The algebra $\bb{R}\otimes\bb{C}\otimes\bb{H}\otimes\bb{O}$ has exactly the right dimensions (32 complex) to describe one generation of fermions. In a closely related approach, Furey studies the minimal ideals of the Clifford algebras $\bb{C}\ell(4)$, and $\bb{C}\ell(6)$, generated from $\bb{C}\otimes\bb{H}$, and $\bb{C}\otimes\bb{O}$ respectively \cite{furey2016standard,furey20183}. In her approach, the leptons and quarks correspond to elements of these minimal ideals, and the gauge symmetries are those unitary symmetries that preserve the ideals. In particular, the part $\bb{C}\otimes\bb{O}$ part of Dixon's algebra can be associated to the color and electric charge internal degrees of freedom, with the color gauge group $SU(3)$ corresponding to the maximal compact subgroup of the exceptional group $G_2$ of automorphisms of the $\bb{O}$ which ﬁxes one of the octonion units.

Many others have contributed to these, and related, algebraic approaches including those based on topology \cite{gresnigt2018braids,gresnigt2019braided,gresnigt2020topological,gresnigt2021topological,gresnigt2021braided}, exceptional Lie groups \cite{manogue2010octonions,wilson2022octonionic,manogue2022octions,wilson2022chirality,raj2022lagrangian,kaushik2022e_8}, Clifford algebras \cite{trayling1999geometric,trayling2001geometric,trayling2004cl,stoica2017standard,stoica2020chiral,todorov2022octonionic, perelman2019c,perelman2019r,pavvsivc2013geometric}, and Jordan algebras \cite{dubois2016exceptional,dubois2019exceptional,todorov2018octonions,todorov2018deducing,boyle2020standard,boyle2020standard2}. 

Existing division algebraic models offer an elegant algebraic construction for the internal space of a single generation of leptons and quarks. Despite several attempts \cite{dixon2013division,furey2014generations,gillard2019three}, a clear algebraic origin for the existence of three generation is yet to be found. The Pati-Salam model, as well as both the $SU(5)$ and $Spin(10)$ grand unified theories likewise correspond to single generation models, lacking any theoretical basis for three generations, which ultimately has to be imposed by hand.

Furey identifies three generations of color states directly from the algebra $\bb{C}\ell(6)$ generated from the adjoint actions of $\bb{C}\otimes\bb{O}$ \cite{furey2014generations}. The algebra $\bb{C}\ell(6)$ is 64 complex dimensional. Constructing two representations of the Lie algebra $su(3)$ within this algebra, the remaining 48 degrees of freedom transform under the action of the $SU(3)$ as three generations of leptons and quarks. The most obvious extension to include $U(1)_{em}$ via the number operator, which works in the context of a one-generation model, fails to assign the correct electric charges to states. A generalized action that leads to a generator that produces the correct electric charges for all states is introduced in \cite{furey2018three}.

Dixon on the other hand considers the algebra $\bb{T}^6=\bb{C}\otimes\bb{H}^2\otimes\bb{O}^3$, where $\bb{T}=\bb{R}\otimes\bb{C}\otimes\bb{H}\otimes\bb{O}$ in order to represent three generations, with a single generation being described by $\bb{T}^2$, a complexified (hyper) spinor in 1+9D spacetime \cite{dixon2004division}. However, the choice $\bb{T}^6$, as opposed to any other $\bb{T}^{2n}$ appears rather arbitrary, although can be motivated from the Leech lattice.

These division algebraic models share many similarities with those based on the exceptional Jordan algebra $J_3(\bb{O})$ consisting of three by three matrices over $\bb{O}$, which has likewise been proposed to describe three generations \cite{dubois2016exceptional,dubois2019exceptional,todorov2018octonions,todorov2018deducing,boyle2020standard2,boyle2020standard}. In these models, each of the three octonions in $J_3(\bb{O})$ is likewise associated with one generation via the three canonical $J_2(\bb{O})$ subalgebras of $J_3(\bb{O})$. 

In \cite{perelman2019r} it is argued that $\bb{R}\otimes\bb{C}\otimes\bb{H}\otimes\bb{O}$-valued gravity can naturally describe a grand unified field theory of Einstein's gravity with a Yang-Mills theory containing the SM, leading to a $SU(4)^4$ symmetry group that potentially extends the SM with an extra fourth family of fermions. The existence of a fourth generation of fermions lacks experimental support however.  

In \cite{manogue1999dimensional} it is shown how, by choosing a privileged $\bb{C}$ subalgebra of $\bb{O}$, it is possible to reduce ten dimensional spacetime represented by $SL(2,\bb{O})$ to four dimensional spacetime $SL(2,\bb{C})$. This process of dimensional reduction naturally isolates three $\bb{H}$ subalgebras of $\bb{O}$: those that contain the privileged $\bb{C}$ subalgebra. These three intersecting $\bb{H}$ subalgebras are subsequently interpreted as describing three generations of leptons.

Starting with $\mathbb{R}$, each of the remaining three division algebras can be generated via what is called the Cayley-Dickson (CD) process. This process does not terminate with $\bb{O}$ however, but continues indefinitely to produce a series of $2^n$-dimensional algebras. We therefore ask the question: Can we go beyond the division algebras, to the CD algebra of sedenions $\bb{S}$, generated from $\bb{O}$, in order to describe three generations?

The present paper advocates that the CD algebra of sedenions $\bb{S}$ constitutes a natural mathematical object which exhibits the algebraic structure necessary to describe the internal space of three generations. We restrict ourselves for the time being to considering only the $SU(3)_C$ color symmetry of leptons and quarks. 

The algebra $\bb{S}$ was first proposed to play a role in describing three generations  in \cite{gillard2019three}. The key idea behind that proposal was to generalize the constructing of three generations of leptons in terms of three $\bb{H}$ subalgebras of $\bb{O}$ in \cite{manogue1999dimensional} to three generations in terms of three $\bb{O}$ subalgebras of $\bb{S}$, where each generation is associated with one copy of $\bb{O}$. One finds, as in \cite{manogue1999dimensional}, that the resulting three generations are not linearly independent. It was suggested in \cite{gillard2019three} (and later in \cite{manogue2022octions}) that this overlap could provide an algebraic basis for neutrino oscillations and quark mixing, although the viability of this idea remains to be investigated.

The model in \cite{gillard2019three} suffers from two significant drawbacks. Each generation comes with its own copy of $SU(3)_C$ thereby also requiring three generations of gluons, for which there is currently no experimental evidence. Additionally, $Aut(\bb{S})=Aut(\bb{O})\times S_3$, where $Aut(\bb{O})=G_2$, and $S_3$ is the permutation group of three objects \cite{schafer1954algebras,brown1967generalized}. The $S_3$ automorphisms of $\bb{S}$ were however not given any clear physical interpretation, in part because these automorphisms stabilize the octonion subalgebras in $\bb{S}$.


The model we presented here builds on \cite{gillard2019three} and seeks to resolve the shortcoming just mentioned.  Instead of associating each $\bb{O}$ subalgebra of $\bb{S}$ with one generation, we use all three $\bb{O}$ subalgebras to construct a single generation. This corresponds to a direct generalization of the construction in \cite{furey2016standard} where three $\bb{H}$ subalgebras of $\bb{O}$ are used to construct a single generation. Subsequently, we utilize sedenion $S_3$ automorphism of order three to generate the additional two generations. This construction provides a clear interpretation of the new $S_3$ automorphism. Furthermore, all three generations transform as required under a single copy of the gauge group $SU(3)_C$, thereby avoiding introducing three generations of gluons.


In the next section we provide a brief overview of the normed division algebras, in particular the quaternions $\bb{H}$ and octonions $\bb{O}$. In Section \ref{Furey} we review the construction of one generation of fermions with unbroken $SU(3)_c\times U(1)_{em}$ gauge symmetry from $\bb{C}\otimes\bb{O}$, following closely \cite{furey2016standard}. The Cayley-Dickson construction and the algebra of sedenions are discussed in Section \ref{sedenions}. Finally we present our three generation model based on the algebra of sedenions in Section \ref{sedenionmodel}. We conclude with an outlook of how to develop the model further, and a discussion.

\section{Normed division algebras}\label{sec:normeddivisionalgebras}

A division algebra is an algebra over a field where division is always well-defined, except by zero. A normed division algebra has the additional property that it is also normed vector spaces, with the norm defined in terms of a conjugate. A well-known result by Hurwitz \cite{hurwitz1898ueber} is that there exist only four normed division algebras (over the field of real numbers): $\bb{R}$, $\bb{C}$, $\bb{H}$, $\bb{O}$, of dimensions one, two, four and eight respectively. In going to higher-dimensional algebras, successive algebraic properties are lost: $\bb{R}$ is self-conjugate, commutative and associative, $\bb{C}$ is commutative and associative (but no longer self-conjugate), $\bb{H}$ is associative but no longer commutative, and finally $\bb{O}$ is neither commutative nor associative (but alternative).

The quaternions $\bb{H}$ are a generalization of the complex numbers $\bb{C}$ with three mutually anticommuting imaginary units $I,J,K$, satisfying $I^2=J^2=K^2=IJK=-1$, which implies $IJ=K=-JI$, $JK=I=-KJ$, and $KI=J=-IK$. A general quaternion $q$ may then be written as
\begin{eqnarray}
    q=q_01+q_1I+q_2J+q_3K,\qquad q_0,q_1,q_2,q_3\in\bb{R},
\end{eqnarray}
With the quaternion conjugate $\overline{q}$ defined as $\overline{q}=q_0-q_1I-q_2J-q_3K$. The norm of a quaternion $\vert q\vert$ is subsequently defines by $\vert q\vert^2=q\overline{q}=\overline{q}q$, and the inverse $q^{-1}=\overline{q}/\vert q\vert^2$. 

The automorphism group of $\bb{H}$ is $SU(2)$. Indeed, there is an isomorphism between the quaternions $\bb{H}$ and the real Clifford algebra $\Cl(0,2)$, while the complexified quaternions $\bb{C}\otimes\bb{H}$ (isomorphic to the Pauli algebra) are isomorphic to the complex Clifford algebra $\CCl(2)$. Note, however, that $\bb{C}\otimes\bb{H}$ is not a division algebra (but remains associative), and manifestly contains projectors, for example: $(1+iK)(1-iK)=0$.


The octonions $\bb{O}$ are the largest division algebra, of dimension eight. Its orthonormal basis comprises seven imaginary units: $i_1,...i_7$, along with the unit $1=i_0$. A general octonion $x$ may then be written as
\begin{eqnarray}
    x=x_0i_0+x_1i_i+...+x_7i_7,\qquad x_0,...,x_7\in\bb{R},
\end{eqnarray}
with the octonion conjugate $\overline{x}$ defined as $\overline{x}=x_0i_0-x_1i_1-...-x_7i_7$. The norm of an octonion $\vert x\vert$ is subsequently defines by $\vert x\vert^2=x\overline{x}=\overline{x}x$, and the inverse $x^{-1}=\overline{x}/\vert x\vert^2$. 

\begin{figure}[h!]
\centering
\includegraphics[scale=0.08]{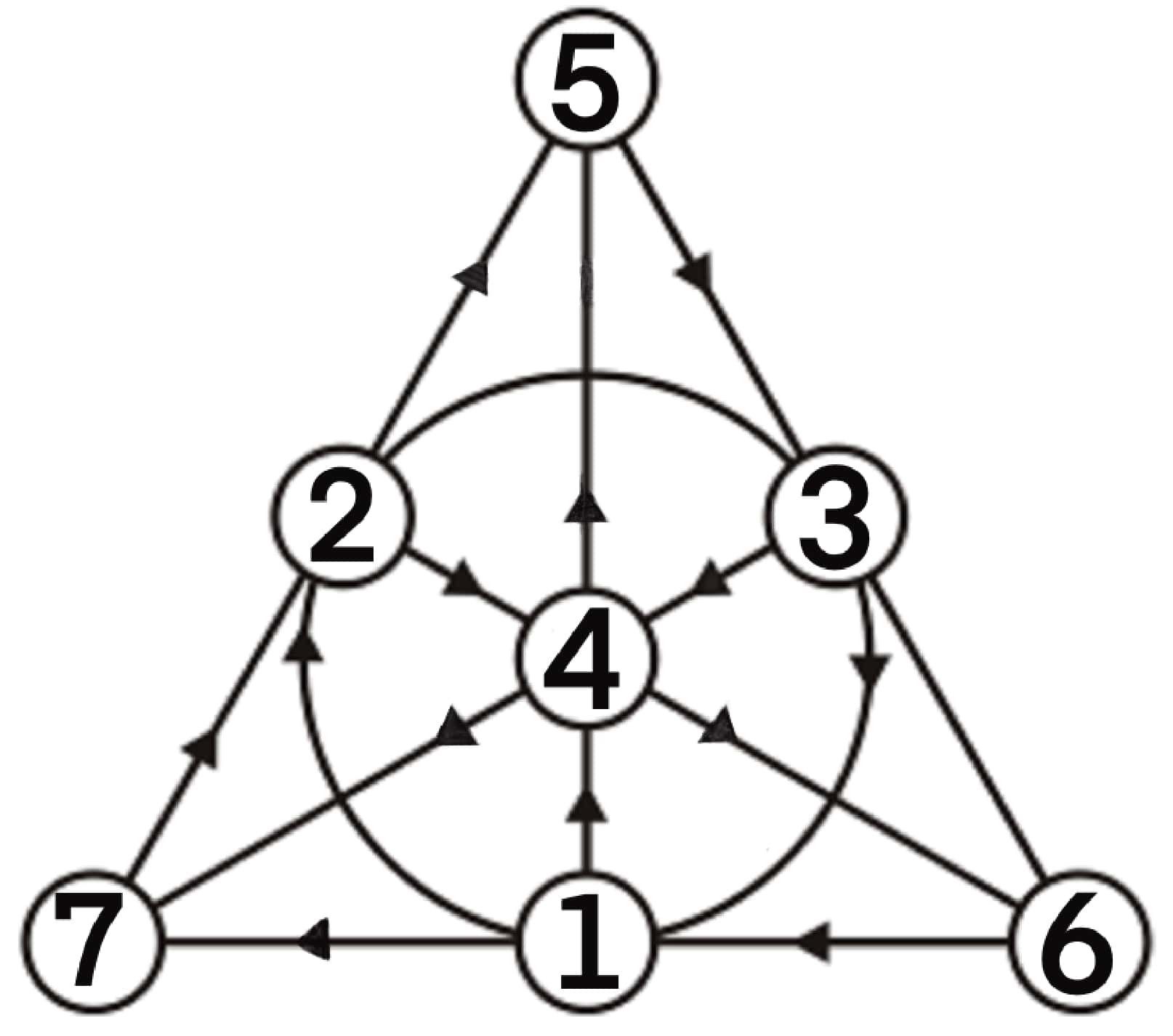}
\caption{The Fano plane, encoding the multiplicative structure of our octonions, where $a\equiv i_a,\; a=1,...,7$. Note that each line is cyclic, representing a quaternionic triple.}
\label{fanoplane}
\end{figure}

The multiplication of octonions\footnote{There are different multiplication rules for $\bb{O}$ used by different authors in the literature. Here we follow the multiplication table used in \cite{lohmus1994nonassociative}} is captured in terms of the Fano plane Fig. \ref{fanoplane}. Each projective line in the Fano plane corresponds (together with the identity $i_0$) to an $\bb{H}$ subalgebra; there are seven such subalgebras. Like with $\bb{H}$, all the imaginary units anticommute under multiplication. Unlike with $\bb{H}$, the multiplication of elements not belonging to the same $\bb{H}$ subalgebra is non-associative. For example $i_4(i_7i_6) = -i_5 \neq i_5 = (i_4i_7)i_6$. Octonion multiplication however is alternative $x(xy) = (xx)y$ and $y(xx) = (yx)x$, $\forall x,y\in\bb{O}$. The complexified octonions $\bb{C}\otimes \bb{O}$ are again not a division algebra (but remains alternative).
\vspace{5pt}

As vector spaces $\bb{O}=\bb{C}^4$. The splitting of $\bb{O}$ as $\bb{C}\oplus\bb{C}^3$ relies on choosing a preferred octonion unit $i_a$ (and hence a preferred $\bb{C}$ subalgebra in $\bb{O}$). For our purpose we choose $i_4$. The map \cite{manogue1999dimensional}
\begin{eqnarray}
\pi(x)=\frac{1}{2}(x+i_4x\Bar{i_4}),\quad x\in\bb{O},
\end{eqnarray}
where $\Bar{i_4}$ indicates the octonion conjugation, then projects $\bb{O}$ down to this preferred $\bb{C}\subset \bb{O}$, and we can write the octonion $x=x_0i_0+...+x_7i_7$ as:
\begin{eqnarray}
x=(x_0+x_4i_4)i_0+(x_1-x_5i_4)i_1+(x_2-x_6i_4)i_2+(x_3-x_7i_4)i_3.
\end{eqnarray}
Note that the product $i_4x\Bar{i_4}$ is defined unambiguously since $\bb{O}$ is alternative. 

The automorphism group of $\bb{O}$ is the $14$-dimensional exceptional Lie group $G_2$. This exceptional group contains $SU(3)$ as one of its maximal subgroups, corresponding to the stabilizer subgroup of one of the octonion imaginary units, or equivalently, the subgroup of $Aut(\bb{O})$ that preserves the representation of $\bb{O}$ as the complex space $\bb{C}\oplus \bb{C}^3$. This splitting is associated with the quark-lepton symmetry \cite{manogue1999dimensional}. The space of internal states of a quark is then the three complex dimensional space $\bb{C}^3$ whereas the internal space of a lepton is $\bb{C}$. 



Since $\bb{O}$ and $\bb{C}\otimes\bb{O}$ are nonassociative, they are not representable as matrix algebras (with the standard matrix product). The algebra generated from the composition of left and right actions of $\bb{O}$ (and $\bb{C}\otimes\bb{O}$) however is associative, since each such left (right) action corresponds to a linear operator (endomorphism). 

Let $L_a$ ($R_a$) denote the linear operator of left (right) multiplication by $a\in\bb{C}\otimes\bb{O}$:
\begin{eqnarray}
    L_a[x]=ax,\quad R_a[x]=xa,\quad \forall a,x\in  \bb{C}\otimes\bb{O}.
\end{eqnarray}
Then
\begin{eqnarray}
    L_aL_b[x]&=&a(bx) \quad \neq \quad L_{ab}[x]=(ab)x,\\
    R_aR_b[x]&=&(xb)a \quad \neq \quad R_{ab}[x]=x(ab).
\end{eqnarray}
The mappings $a\rightarrow L_a$ and $a\rightarrow R_a$ do not correspond to algebra homomorphisms as they each generate an associative algebra called the \textit{associative multiplication algebra}\footnote{We will henceforth refer to this algebra simply as the left action algebra or left multiplication algebra}. They do however preserve the quadratic relations $\langle x,y\rangle=\frac{1}{2}(x\overline{y}+y\overline{x})$ where $x,y\in \bb{C}\otimes\bb{O}$
\begin{eqnarray}
    L_xL_{\overline{y}}1+L_yL_{\overline{x}}1=2\langle x,y \rangle 1=R_xR_{\overline{y}}1+R_yR_{\overline{x}}1.
\end{eqnarray}

Since $L_a$ ($R_a$) correspond to linear operators, they can be represented as $8\times 8$ complex matrices (acting on the vector space $\bb{C}\otimes\bb{O}$ written as a column vector).

For $\bb{C}\otimes\bb{O}$, one finds the following identities \cite{dixon2013division,furey2016standard}:
\begin{eqnarray}
    L_{i_1}L_{i_2}L_{i_3}L_{i_4}L_{i_5}L_{i_6}x&=&L_{i_7}x,\\
    ...L_{i_b}L_{i_a}L_{i_a}L_{i_c}....x&=&-...L_{i_b}L_{i_c}....x,\\
    ...L_{i_a}L_{i_b}...x&=&-...L_{i_b}L_{i_a}...,
\end{eqnarray}
where $a,b,c=1,...,7$.

Due to the nonassociativity of $\bb{O}$, the left (right) associative multiplication algebra of $\bb{C}\otimes\bb{O}$ contains genuinely new maps which are not captured by $\bb{C}\otimes\bb{O}$. For example, $i_3(i_4(i_6+i_2))\neq y(i_6+i_2)$ for any $y\in\bb{C}\otimes\bb{O}$. 
There are a total of 64 distinct left-acting complex-linear maps from $\bb{C}\otimes\bb{O}$ to itself, and these (due to the given identities above) provide a faithful representation of $\CCl(6)$. 

Denoting the 64-dimensional left (right) associative multiplication algebra generated from left (right) actions of $\bb{C}\otimes\bb{O}$ on itself by $(\bb{C}\otimes\bb{O})_L$ ($(\bb{C}\otimes\bb{O})_R$), one finds that any left (right) action can always be rewritten as a right (left) action \cite{dixon2013division}. That is: 
\begin{eqnarray}
(\bb{C}\otimes\mathbb{O})_L\cong (\bb{C}\otimes\mathbb{O})_R\cong \bb{C}\ell(6).
\end{eqnarray}
This is in contrast to a similar construction for $\bb{C}\otimes\bb{H}$, where one finds that the left and right actions are genuinely distinct, each generating a copy of $\bb{C}\ell(2)$. The left and right adjoint actions in this case commute, and only by considering both does one obtain a basis for $Mat(4,\bb{C})\cong\bb{C}\ell(4)$.


\section{One generation of electrocolor states from $\bb{C}\otimes\bb{O}$}\label{Furey}

Let $e_1:= L_{i_1},...,e_6:=L_{i_6}$ be a generating basis (over $\bb{C}^6$) for $\bb{C}\ell(6)$ associated with the left multiplication algebra of the complex octonions, satisfying $e_i^2=-1,\;e_ie_j=-e_je_i$. Define the Witt basis
\begin{eqnarray}\notag
\alpha_1^\dagger &=& \frac{1}{2}(e_1+ie_5),\quad \alpha_2^\dagger = \frac{1}{2}(e_2+ie_6),\quad \alpha_3^\dagger = \frac{1}{2}(e_3+ie_7)\\\notag
\alpha_1 &=& \frac{1}{2}(-e_1+ie_5), \quad \alpha_2 = \frac{1}{2}(-e_2+ie_6),\quad \alpha_3 = \frac{1}{2}(-e_3+ie_7).
\end{eqnarray}
Here ${}^{\dagger}$ corresponds to the composition of complex and octonion conjugation. This new basis satisfies the anticommutation relations \footnote{Note that as left actions we have $\{L_{\alpha_i},L_{\alpha_j}\}x=0$ etc for $x\in\bb{C}\otimes\bb{O}$.}
\begin{eqnarray}
    \{\alpha_i,\alpha_j\}=\{\alpha_i^{\dagger},\alpha_j^{\dagger}\}=0, \quad\{\alpha_i,\alpha_j^{\dagger}\}=\delta_{ij}
\end{eqnarray}

Each pair of ladder operators in isolation generates $\bb{C}\ell(2)$, and is associated with one of the three $\bb{H}$ subalgebras of $\bb{O}$ that contain the privileged complex subalgebra generated by $i_4$. Subsequently, we obtain three anticommuting copies of $\bb{C}\ell(2)$, which when considered together generate the full $\bb{C}\ell(6)$ left multiplication algebra of $\bb{C}\otimes\bb{O}$.

From the Witt basis, it is possible to construct two minimal left ideals of the algebra $\bb{C}\ell(6)$, following the procedure in \cite{ablamowicz1995construction} (see \cite{furey2016standard} for a detailed construction):
\begin{eqnarray}
\bb{C}\ell(6)\omega\omega^{\dagger},\qquad \bb{C}\ell(6)\omega^{\dagger}\omega,
\end{eqnarray}
where $\omega = \alpha_1\alpha_2\alpha_3$ and $\omega^\dagger = \alpha_3^\dagger\alpha_2^\dagger\alpha_1^\dagger$ are nilpotents, but $\omega\omega^{\dagger}$ and $\omega^{\dagger}\omega$ are primitive idempotents. Each ideal is eight complex dimensional. Explicitly,
\begin{equation} \notag
 \begin{split}
     S^u=&{\nu}\omega\omega^\dagger + \\  \overline{d^r}{\alpha_1^\dagger}\omega\omega^\dagger + &\overline{d^g}{\alpha_2^\dagger}\omega\omega^\dagger + \overline{d^b}{\alpha_3^\dagger}\omega\omega^\dagger + \\ u^r{\alpha_3^\dagger}{\alpha_2^\dagger}\omega\omega^\dagger + &u^g{\alpha_1^\dagger}{\alpha_3^\dagger}\omega\omega^\dagger + u^b{\alpha_2^\dagger}{\alpha_1^\dagger}\omega\omega^\dagger + \\ &e^+{\alpha_3^\dagger}{\alpha_2^\dagger}{\alpha_1^\dagger}\omega\omega^\dagger
\end{split}
\qquad 
\begin{split}
     S^d=&{\overline{\nu}}\omega^\dagger\omega + \\ {d}^r{\alpha_1}\omega^\dagger\omega + &{d}^g{\alpha_2}\omega^\dagger\omega +  {d}^b{\alpha_3}\omega^\dagger\omega + \\ \overline{u^r}{\alpha_3}{\alpha_2}\omega^\dagger\omega + &\overline{u^g}{\alpha_1}{\alpha_3}\omega^\dagger\omega + \overline{u^b}{\alpha_2}{\alpha_1}\omega^\dagger\omega + \\ &e^-{\alpha_3}{\alpha_2}{\alpha_1}\omega^\dagger\omega
\end{split}
\end{equation}
where the suggestively labelled coefficients are elements of $\bb{C}$. 

The unitary symmetries that preserve the Witt basis, and hence the minimal left ideals is $U(3)=SU(3)\times U(1)$. The generators of this symmetry, written in terms of the Witt basis, are:
    \begin{align} \notag
     \Lambda_1 &= -\alpha_2^\dagger\alpha_1 - \alpha_1^\dagger\alpha_2 \notag \hspace{2em}
     \Lambda_2 = i\alpha_2^\dagger\alpha_1 -i\alpha_1^\dagger\alpha_2 \notag \hspace{2em}
     \Lambda_3 = \alpha_2^\dagger\alpha_2 - \alpha_1^\dagger\alpha_1 \notag\\ 
     \Lambda_4 &= -\alpha_1^\dagger\alpha_3 - \alpha_3^\dagger\alpha_1 \notag \hspace{2em}
     \Lambda_5 = -i\alpha_1^\dagger\alpha_3 + i\alpha_3^\dagger\alpha_1 \notag \hspace{2em}
     \Lambda_6 = -\alpha_3^\dagger\alpha_2 - \alpha_2^\dagger\alpha_3 \notag\\
     \Lambda_7 &= i\alpha_3^\dagger\alpha_2 - i\alpha_2^\dagger\alpha_3 \notag \hspace{1em}
     \Lambda_8 = -\frac{1}{\sqrt{3}}(\alpha_1^\dagger\alpha_1 + \alpha_2^\dagger\alpha_2 - 2\alpha_3^\dagger\alpha_3), \notag
 \end{align}
     \begin{equation} \notag
     Q = \frac{1}{3}(\alpha_1^\dagger\alpha_1 + \alpha_2^\dagger\alpha_2 + \alpha_3^\dagger\alpha_3).
 \end{equation}
The basis states of minimal ideals transform as $1\oplus 3\oplus\overline{3}\oplus1$ under $SU(3)$, and this symmetry can therefore be associated with the color symmetry $SU(3)_C$, justifying the choice of coefficients. The $U(1)$ generator $Q$, related to the number operator $Q=N/3$, on the other hand gives correct electric charge for each state. The ideal $S^u$ contains the isospin up states, whereas the $S^d$ contains the isospin down states.

One generation of leptons and quarks with correct unbroken $SU(3)_C\times U(1)_{em}$ symmetry can therefore be elegantly represented in terms of two minimal left ideals of $\bb{C}\ell(6)$ generated from $\bb{C}\otimes\bb{O}$. The dimension of the minimal ideals dictates the number of distinct physical states, whereas the gauge symmetries are those unitary symmetries that preserve the ideals (or equivalently, the Witt basis).

\section{The Cayley-Dickson construction and the algebra of sedenions}\label{sedenions}


The CD process is an iterative construction that generates at each stage an algebra (with involution) of dimension twice that of the previous. Each algebra is constructed as a direct sum of the previous algebra, so that $\bb{C}=\bb{R}\oplus\bb{R}i$ where $i$ is the complex structure introduced in the process. Similarly, $\bb{H} = \bb{C} \oplus \bb{C} J$, where $i$, $J$ and $iJ$ are identified with the quaternion imaginary bases $I, J, K$, and  similarly $\bb{O} = \bb{H}\oplus\bb{H}i_4$. 

This process does not terminate with $\bb{O}$ but continues, generating a series of $2^n$-dimensional (non division) algebras. A generic element of the CD algebra $\bb{A}_n$, can then  be written as $a+bu$, where $a,b \in \bb{A}_{n-1}$, and $u$ is the new imaginary unit introduced by the CD process applied to $\bb{A}_{n-1}$. The fifth CD algebra $\bb{A}_{4}$ ($\bb{A}_{0}=\bb{R}$), generated from $\bb{O}$, is the 16-dimensional algebra of sedenions $\bb{S}$. This algebra is non-commutative, non-associative, and not even alternative ($x(xy) \neq (xx)y$ and $y(xx) \neq (yx)x$ in general). Other properties, like flexibility ($(xy)x = x(yx)$) and power-associativity ($x^n$ associative), still hold (and hold for all CD algebras). 

An orthonormal basis for $\mathbb{S}$ comprises 15 mutually anticommuting imaginary units $s_1,...,s_{15}$ together with the unit $1=s_0$. The imaginary units $s_1,..,s_7$ correspond to the original octonion units $i_1,...,i_7$. A general sedenion may then be written as
\begin{eqnarray}
    w=w_0s_0+w_1s_1+...+w_{15}s_{15},\quad w_0,...,w_{15}\in\bb{R}.
\end{eqnarray}
with the sedenion conjugate $\overline{w}$ defined as $\overline{w}=w_0s_0-w_1s_1-...-w_{15}s_{15}$, and the sedenion norm $\vert w\vert$ defined by $\vert w\vert ^2=w\overline{w}=\overline{w}w$. Whenever $\vert w\vert ^2\neq 0$, the inverse of $w$ is given by $w^{-1}=\overline{w}/\vert w \vert^2$.

The product of two sedenions $w,v$ can be determined using the multiplication table of the orthonormal basis units of $\bb{S}$ provided in Appendix A, source \cite{lohmus1994nonassociative,cawagas2004structure}, together with linearity.

Because $\bb{S}$ is not a division algebra, it contains zero divisors. These are elements of the form
\begin{eqnarray}
(s_a+s_b)(s_c+s_b)=0,\qquad s_a,s_b,s_c,s_d\in\mathbb{S}.
\end{eqnarray}
There are 84 such zero divisors, and the subspace of zero divisors of unit norm is homeomorphic to $G_2$ \cite{moreno1997zero}. 

\subsection{Octonion subalgebras inside the sedenions}

Let us now use $\{e_0,e_1,...,e_{n^2-1}\}$ to denote an orthonormal basis for $\bb{A}_n$, so that
\begin{eqnarray}
    \bb{A}_1=\bb{C}&:& \{e_0,e_1\},\quad\textrm{where}\quad e_1=i,\\
    \bb{A}_2=\bb{H}&:& \{e_0,e_1,e_2,e_3\},\quad\textrm{where}\quad e_1=i=I,\; e_2=J,\; e_3=K\\
    \bb{A}_3=\bb{O}&:& \{e_0,e_1,...,e_7\},\quad\textrm{where}\quad e_1=i=I=i_1,...,e_7=i_7\\
    \bb{A}_4=\bb{S}&:& \{e_0,e_1,...,e_{15}\},\quad\textrm{where}\quad e_1=i=I=i_1=s_1,...,e_{15}=s_{15}.
\end{eqnarray}

Consider $\bb{A}_2=\bb{H}$ with basis $\{e_0,e_1,e_2,e_3\}$. There are three subalgebras isomorphic to $\bb{C}$ within $\bb{H}$, each containing the identity and one of the imaginary units of $\bb{H}$. These subalgebras correspond to three different complex structures in $\bb{H}$, and the common intersection of these three $\bb{C}$ subalgebras is isomorphic to $\bb{R}$. The automorphism group of $\bb{H}$ is $Aut(H)=SU(2)$. The subset of automorphisms of $\bb{H}$ that preserve a given complex structure is $U(1)$, corresponding to the element wise stabilizer subgroup of $SU(2)$.

Applying the CD process to $\bb{H}$ generates $\bb{O}$ with basis $\{e_0,e_1,e_2,e_3,e_4,e_5,e_6,e_7\}$ where $e_4$ is the newly introduced anticommuting imaginary unit and $e_ie_4=e_{i+4}$. Via this same construction, each of the three $\bb{C}\subset\bb{H}$ subalgebras generate a quaternion:
\begin{eqnarray}
\bb{C}_1: \{e_0,e_1\}\xrightarrow{CD} \bb{H}_1: \{e_0,e_1,e_4,e_5\},\\
\bb{C}_2: \{e_0,e_2\}\xrightarrow{CD} \bb{H}_2: \{e_0,e_2,e_4,e_6\},\\
\bb{C}_3: \{e_0,e_3\}\xrightarrow{CD} \bb{H}_3: \{e_0,e_3,e_4,e_7\}.
\end{eqnarray}
The common intersection of these three $\bb{H}\in\bb{O}$ is isomorphic to $\bb{C}$, spanned by $e_0$ and $e_4$. These $\bb{H}_i$, $i=1,2,3$ however are not the only $\bb{H}$ subalgebras of $\bb{O}$, and in total there are seven such subalgebras. However $\bb{H}_1, \bb{H}_2, \bb{H}_3$ are the only (quaternions) subalgebras that contain the new $e_4$. Together with the identity, $e_4$ corresponds to a complex structure.


Applying the CD process to $\bb{O}$ generates $\bb{S}$. Apart from the original (\textit{principal}) $\bb{O}$ with basis $\{e_0,e_1,...,e_7\}$, $\bb{S}$ contains seven $\bb{O}$ subalgebras (which we call \textit{non-principal}). These are explicitly listed in \cite{cawagas2004structure}. Interestingly, all of these contain the new imaginary unit $e_8$, introduced in the CD construction of $\bb{S}$ from $\bb{O}$\footnote{Note that this is a new feature that appears with $\bb{S}$. There are three new, (that is excluding the original $\bb{H}$ with basis $\{e_0,e_1,e_2,e_3\}$ used to generate $\bb{O}$) $\bb{H}$ subalgebras of $\bb{O}$ that do no contain $e_4$. For $\bb{S}$ there are no new $\bb{O}$ that do not contain $e_8$.}. Via this process, each of $\bb{H}_i\subset\bb{O}$ above generates one $\bb{O}_i\in\bb{S}$.  
\begin{eqnarray}
\bb{H}_1: \{e_0,e_1,e_4,e_5\}\xrightarrow{CD} \bb{O}_1: \{e_0,e_1,e_4,e_5,e_8,e_9,e_{12},e_{13}\},\\
\bb{H}_2: \{e_0,e_2,e_4,e_6\}\xrightarrow{CD} \bb{O}_2: \{e_0,e_2,e_4,e_6,e_8,e_{10},e_{12},e_{14}\},\\
\bb{H}_3: \{e_0,e_3,e_4,e_7\}\xrightarrow{CD} \bb{O}_3: \{e_0,e_3,e_4,e_7,e_8,e_{11},e_{12},e_{15}\}.
\end{eqnarray}
Since each $\bb{H}_i$ contains $e_4$, and each $\bb{O}_i$ contains $e_8$, it follows that each $\bb{O}_i$ also contains $e_4e_8=e_{12}$. That is, common intersection of the three $\bb{O}_i$ now corresponds to a $\bb{H}\subset\bb{S}$:
\begin{eqnarray}\label{commonintersection}
\bb{O}_1\cap \bb{O}_2\cap\bb{O}_3=\bb{H},
\end{eqnarray}
where this $\bb{H}$ is generated by $\{e_0,e_4,e_8,e_{12}\}$. $\bb{O}_1, \bb{O}_2, \bb{O}_3$ are the only (octonion) subalgebras of $\bb{S}$ that contain $e_4,e_8$, and $e_{12}$. Together with the identity, this corresponds to a quaternionic structure. 

In addition to the eight octonion subalgebras of $\bb{S}$, there are also a further seven quasi-octonion subloops $\tilde{\bb{O}}$, satisfying all the same properties of the octonion subalgebras, except for the Moufang identities\footnote{The Moufang identities, describing a weaker notion of associativity, are essentially equivalent to left and right alternativity, and flexibility}. As such, they are not isomorphic to the octonion subalgebras. None of the $\tilde{\bb{O}}$ contain the element $e_8$.


\subsection{The left multiplication algebra of $\bb{C}\otimes\bb{S}$}

Despite the algebra $\bb{S}$ being non-associative and non-alternative, just as for $\bb{O}$, we can consider the left actions of $\bb{S}$ on itself as linear operators generating an associative algebra. 

The generalisation from $(\bb{C}\otimes\bb{O})_L$ to $(\bb{C}\otimes\bb{S})_L$ is not immediately obvious because the identities which held for $(\bb{C}\otimes\bb{O})_L$, namely
\begin{eqnarray}
    L_{i_1}L_{i_2}L_{i_3}L_{i_4}L_{i_5}L_{i_6}x&=&L_{i_7}x,\\
    ...L_{i_b}L_{i_a}L_{i_a}L_{i_c}....x&=&-...L_{i_b}L_{i_c}....x,\label{octidentity1}\\
    ...L_{i_a}L_{i_b}...x&=&-...L_{i_b}L_{i_a}...,\label{octidentity2}
\end{eqnarray}
where $a,b,c=1,...,7$ are no longer satisfied by general composition of sedenion left multiplications. Although one finds a new identify
\begin{align}
\label{sedenionidentity}
L_{s_1}L_{s_2}...L_{s_{14}}w=L_{s_{15}}w,
\end{align}
the two identities (\ref{octidentity1}) and (\ref{octidentity2}), crucial for generating a Clifford algebra, no longer hold in general. For example, $s_3(s_{14}(s_1)) = -s_{14}(s_3(s_1))$, but $s_3(s_{14}(s_2)) = +s_{14}(s_3(s_2)$. 
Consequently, the left multiplications of the (complex) sedenions do not generate $\bb{C}\ell(14)$, as one might initially expect.

However, since the linear operators corresponding to each complex sedenion left multiplication can be written as a $16\times 16$ matrix with complex entries acting on $\bb{C}\otimes\bb{S}$ written as a column vector, one would expect to be able to generate $Mat(16,\bb{C})\cong \bb{C}\ell(8)$. It then remains to find a suitable set of sedenion elements that generate $\bb{C}\ell(8)$ via their left multiplication. Closer inspection reveals that all the left multiplications of the original octonion elements $e_0=i_0=s_0,...,e_7=i_7=s_7$ do satisfy the identities (\ref{octidentity1}) and (\ref{octidentity2}) (where now the action is on $\bb{C}\otimes\bb{S}$ instead of $\bb{C}\otimes\bb{O}$), 
as do all new sedenion basis elements $e_9=s_7,...,e_{15}=s_{15}$:
\begin{eqnarray}
    e_i(e_jw)&=&-e_j(e_iw),\qquad e_{i+8}(e_{j+8}w)= -e_{j+8}(e_{i+8}w),\\
    e_i(e_iw)&=&-w,\qquad e_{i+8}(e_{i+8}w)= -w\quad i,j=1,...,7,\quad \forall w\in\bb{C}\otimes\bb{S}
\end{eqnarray}
However, $e_i$ and $e_{i+8}$ fail to anti-commute (or commute) with each other as left actions
\begin{eqnarray}
   e_i(e_{j+8}w)\neq -e_{j+8}(e_iw). 
\end{eqnarray}
The left action of $e_8$ however anti-commutes with the left action of every other basis element. One possible generating basis for $\bb{C}\ell(8)$ is therefore given by the left multiplications of $\{e_1,...,e_8\}$. Another possible generating basis is $\{e_8,...,e_{15}\}$.

Accepting some abuse of notation, we now simple write $L_{e_i}=e_i$ and take ${e_0,e_1,...,e_8}$ as our generating basis for $\bb{C}\ell(8)$. The left action of the remaining sedenion basis elements $e_9,...,e_{15}$ then need to be expressed as the left action of some element of $\bb{C}\ell(8)$. After some trial and error, one finds that
\begin{eqnarray}
    e_9w&=&\frac{1}{2}(-e_{123458}+e_{123678}+e_{145678}-e_{18})w,\\
    e_{10}w&=&\frac{1}{2}(-e_{123468}-e_{123578}+e_{245678}-e_{28})w,\\
    e_{11}w&=&\frac{1}{2}(-e_{123478}+e_{123568}+e_{345678}-e_{38})w\\
    e_{12}w&=&\frac{1}{2}(e_{124568}+e_{134578}+e_{234678}-e_{48})w\\
    e_{13}w&=&\frac{1}{2}(e_{124578}-e_{134568}+e_{235678}-e_{58})w\\
    e_{14}w&=&\frac{1}{2}(e_{124678}-e_{135678}-e_{234568}-e_{68})w\\
    e_{15}w&=&\frac{1}{2}(e_{125678}+e_{134678}-e_{234578}-e_{78})w,
\end{eqnarray}
where $e_{123458}=e_1e_2e_3e_4e_5e_8=L_{e_1}L_{e_2}L_{e_3}L_{e_4}L_{e_5}L_{e_8}$, $e_{18}=e_1e_8=L_{e_1}L_{e_8}$ etc are elements of $\bb{C}\ell(8)$. 
\subsection{Automorphisms of sedenions}

Schafer \cite{schafer1954algebras} showed that for CD algebras $\bb{A}_n$ with $n\geq 4$ ($\bb{A}_0=\bb{R}$), the derivation algebra $\mathfrak{der}(\bb{A}_n)$  consists of derivations of the form $a+bu\rightarrow aD+(bD)u$, where $a,b\in\bb{A}_{n-1}$, $u$ is the new anticommuting imaginary unit $u$ introduces in the CD construction of $\bb{A}_n$ from $\bb{A}_{n-1}$, and $D$ is a derivation of $\bb{A}_{n-1}$. Brown \cite{brown1967generalized} demonstrated that if $\theta\in Aut(\bb{A}_{n-1})$, then 
\begin{eqnarray}
\theta'&:& a+bu\rightarrow a\theta+(b\theta)u,\\
\epsilon&:& a+bu\rightarrow a-bu,\\
\psi&:& a+bu\rightarrow \frac{1}{4}[a+3a^*+\sqrt{3}(b-b^*)]+\frac{1}{4}[b+3b^*-\sqrt{3}(a-a^*)]u,
\end{eqnarray}
are automorphisms of $\bb{A}_n$. Here ${}^*$ denotes conjugation in $\bb{A}_{n-1}$,  and $\epsilon$ and $\psi$, satisfying $\epsilon^2=\psi^3=1,\;\epsilon\psi=\psi^2\epsilon$, generate $S_3$. That is, in general
\begin{eqnarray}
Aut(\bb{A}_{n-1})\times S_3\subseteq Aut(\bb{A}_n).
\end{eqnarray}

It follows that $SU(2)\times S_3$ are automorphisms of $\bb{O}$, but crucially, these are not all of the octonion automorphisms: $SU(2)\times S_3\subset G_2$. However, for the cases where $n=4,5,6$, the equality holds \cite{brown1967generalized}
\begin{eqnarray}
Aut(\bb{A}_{n-1})\times S_3= Aut(\bb{A}_n),\quad n=4,5,6.
\end{eqnarray}
In particular, this means that:
\begin{eqnarray}
Aut(\bb{S})=Aut(\bb{O})\times S_3=G_2\times S_3.
\end{eqnarray}
Explicitly, the automorphisms of $\bb{S}$ are therefore give by
\begin{eqnarray}
\theta'&:& a+be_8\rightarrow a\theta+(b\theta)e_8,\\
\epsilon&:& a+be_8\rightarrow a-be_8,\\
\psi&:& a+be_8\rightarrow \frac{1}{4}[a+3\overline{a}+\sqrt{3}(b-\overline{b})]+\frac{1}{4}[b+3\overline{b}-\sqrt{3}(a-\overline{a})]e_8,
\end{eqnarray}
where $a,b\in\bb{O}$.
The explicit action of $\psi$ on the sedenion basis elements can be written as:
\begin{eqnarray}
       \psi(e_i)&=& -\frac{1}{2}e_i-\frac{\sqrt{3}}{2}e_{i+8},\\
       \psi(e_{i+8})&=& -\frac{1}{2}e_{i+8}+\frac{\sqrt{3}}{2}e_i,\\
       \psi(e_8)&=&e_8
\end{eqnarray}
where $i=1,...,7$. The automorphism $\psi$ corresponds to a simultaneous rotations in the seven $e_i-e_{i+8}$ planes by $2\pi/3$, and therefore does not correspond to an automorphism of $\bb{C}\otimes\bb{O}$. It is also possible to write the $S_3$ automorphisms in matrix form:
\begin{eqnarray}
\psi&:& \begin{pmatrix}
    e_i\\
    e_{i+8}
\end{pmatrix}\rightarrow\begin{pmatrix}
    -1/2 & -\sqrt{3}/2\\
    \sqrt{3}/2 & -1/2
\end{pmatrix}\begin{pmatrix}
    e_i\\
    e_{i+8}
\end{pmatrix}\\
\epsilon&:& \begin{pmatrix}
    e_i\\
    e_{i+8}
\end{pmatrix}\rightarrow\begin{pmatrix}
    1 & 0\\
    0 & -1
\end{pmatrix}\begin{pmatrix}
    e_i\\
    e_{i+8}
\end{pmatrix},\quad i=1,...,7
\end{eqnarray}
The fundamental symmetries of $\bb{S}$ are the same as those of $\bb{O}$, although one find an additional $S_3$ symmetry, suggesting a threefold multiplicity of the automorphisms of $\bb{O}$. 

From the action of $\theta'$ on the sedenion units above it is immediately clear that $e_8$ is stabilized by the $G_2$ automorphisms. Furthermore, these $G_2$ automorphisms map $e_i, i<8$ to $e_j, j<8$, and $e_{i+8}$ to $e_{j+8}$. They therefore do not mix the new sedenion elements $e_{i+8}$ with the original octonion elements $e_i$, $i=1,...,7$. Only the $S_3$ automorphism $psi$ of order three mixes the original octonion units with the new sedenion units. 

Given that the stabilizer of $e_4$ in $G_2$ is $SU(3)$, and $e_8$ is likewise an $SU(3)$ singlet (as it is fixed by $G_2$), it follows that the $SU(3)$ subgroup of $G_2$ fixes the entire quaternion generated by $\{e_0,e_4,e_8,e_4e_8=e_{12}\}$. The $S_3$ automorphisms on the other hand do not fix this quaternion, although they do stabilize it. Note that this quaternion corresponds precisely to the common intersection of our previously isolated octonion subalgebras $\bb{O}_i$, see eqn. (\ref{commonintersection}).

\section{Three generations of color states from $\bb{C}\otimes\bb{S}$}\label{sedenionmodel}

\subsection{Why not sedenions?}
Since one generation of electrocolor states are efficiently represented starting from $\bb{C}\otimes\bb{O}$, one might ask if $\bb{S}$ (or rather $\bb{C}\otimes\bb{S}$) is an appropriate larger algebraic structure capable of describing three generations. Not being a division algebra is not grounds to disqualify $\bb{S}$, for we point out that neither $\bb{C}\otimes\bb{O}$ nor $\bb{C}\ell(6)$ generated as the left multiplication algebra are themselves division algebras. Furthermore, the algebra $\bb{R}\otimes\bb{C}\otimes\bb{H}\otimes\bb{O}$ is, like $\bb{S}$, not even alternative. in fact, the construction of invariant subspaces (minimal ideals) relies explicitly on the use of projectors (and nilpotents), which altogether do not exist in division algebras.

There are several natural reasons to suspect that $\bb{S}$ exhibits the algebraic structure necessary to describe three full generations:
\begin{enumerate}
    \item $Aut(\bb{S})=Aut(\bb{O})\times S_3$, and one finds a threefold multiplicity of the symmetries associated with $\bb{O}$,
    \item The process in Section \ref{sedenions} shows how to naturally isolate thee $\bb{O}$ subalgebras within $\bb{S}$, which could perhaps be used to construct three generations, 
    \item The group $Spin(8)$ generated from $\CCl(8)$ admits a triality, which has on occasion been suggested as a potential source of three generations \cite{boyle2020standard2,masi2021exceptional}. The group of outer-automorphisms of $Spin(8)$ is precisely $S_3$.
\end{enumerate}


One approach to construct three generations with $SU(3)_C$ symmetry is to use each of the three $\bb{C}\otimes\bb{O}_i\subset \bb{S}$ to generate (via its left action on itself, but not as a left action of $\bb{C}\otimes\bb{S}$!) a $\bb{C}\ell(6)$ algebra, and subsequently representing three generations in terms of the minimal left ideals of these three $\bb{C}\ell(6)$. This approach was considered in \cite{gillard2019three}, as a generalization of the construction of three generations of leptons from three $\bb{H}\subset \bb{O}$ developed in \cite{manogue1999dimensional}. 

There are several drawbacks to this approach however.  Each generation requires its own copy of $SU(3)$ resulting in three generations of gluons. Finally, the physical interpretation of the $S_3$ automorphisms remains obscured, because these automorphisms stabilize the (non-principle) octonion subalgebras in $\bb{S}$. 

The approach pursued here is different and seeks to resolve these shortcomings. We instead use each $\bb{C}\otimes\bb{O}_i\subset \bb{S}$ to construct a pair or fermionic ladder operators that each generate $\bb{C}\ell(2)$, via their left multiplication action on all of $\bb{C}\otimes\bb{S}$ (instead of just $\bb{C}\otimes\bb{O}_i$). The three pairs of ladder operators are independent of one another and hence we identify a single copy of $\bb{C}\ell(6)\cong\bb{C}\ell(2)\hat{\otimes}\bb{C}\ell(2)\hat{\otimes}\bb{C}\ell(2)$, corresponding to a subalgebra of $\bb{C}\ell(8)$. Thus, we will employ all three $\bb{O}_i\subset \bb{S}$ in order to construct a single generation of states. This corresponds to a direct generalization of the construction of reviewed in Section \ref{Furey} where three $\bb{H}$ subalgebras of $\bb{O}$ are used to construct a single generation of color states. Subsequently, the order three $S_3$ automorphism of $\bb{S}$ will be used to generate two additional generations. This construction will therefore provide a clear interpretation of the new $S_3$ automorphism $\psi$. All three generations constructed in this manner transform as required under a single copy of the gauge group $SU(3)_C$, thereby avoiding introducing three generations of gluons.

\subsection{One generation of electrocolor states from $\bb{C}\otimes\bb{S}$}\label{onegenS}

We proceed by first constructing a single generation with unbroken $SU(3)_C\times U(1)$ symmetry from $\bb{C}\otimes\bb{S}$. Considering the three octonion subalgebras $\bb{O}_1,\bb{O}_2,\bb{O}_3$ of $\bb{S}$ defined above:
\begin{eqnarray}
\bb{O}_1&:& \{e_0,e_1,e_4,e_5,e_8,e_9,e_{12},e_{13}\}\\
\bb{O}_2&:& \{e_0,e_2,e_4,e_6,e_8,e_{10},e_{12},e_{14}\}\\
\bb{O}_3&:& \{e_0,e_3,e_4,e_7,e_8,e_{11},e_{12},e_{15}\}.
\end{eqnarray}

For each $\bb{O}_i$ subalgebra, we define a single pair of raising and lowering operators as follows:
\begin{eqnarray}
\label{newladder1} A_1^{\dagger}&\equiv&\frac{1}{2\sqrt{2}}(e_1+ie_5+e_9+ie_{13}),\quad A_1\equiv \frac{1}{2\sqrt{2}}(-e_1+ie_5-e_9+ie_{13})\\
\label{newladder2} A_2^{\dagger}&\equiv& \frac{1}{2\sqrt{2}}(e_2+ie_6+e_{10}+ie_{14}),\quad A_2\equiv \frac{1}{2\sqrt{2}}(-e_2+ie_6-e_{10}+ie_{14})\\
\label{newladder3} A_3^{\dagger}&\equiv&\frac{1}{2\sqrt{2}}(e_3+ie_7+e_{11}+ie_{15}),\quad A_3\equiv \frac{1}{2\sqrt{2}}(-e_3+ie_7-e_{11}+ie_{15}).
\end{eqnarray}
It is readily checked that $A_i(A_iw)=A_i^{\dagger}(A_i^{\dagger}w)=0$ and $A_i(A_jw)=-A_j(A_iw)$, $\forall w\in \bb{C}\otimes\bb{S}$, and therefore these ladder operators satisfy (as left actions on a general $w\in\bb{S}$) the usual anticommutation relations:
\begin{eqnarray}
    \{A_i,A_j\}w=\{A_i^{\dagger},A_j^{\dagger}\}w=0,\quad \{A_i,A_j^{\dagger}\}w=\delta_{ij}w,\; \forall w\in \bb{C}\otimes\bb{S}.
\end{eqnarray}
Each $A_i\in \bb{O}_i\subset \bb{S}$ in a generalization of $\alpha_i\in\bb{H}_i\subset \bb{O}$ where now each ladder operators consists of four terms instead of just two.

Subsequently we can proceed to construct two minimal left ideals, in a manner identical as in Section \ref{Furey}. These ideals are identical to $S^u$ and $S^d$ above, and (as will be demonstrated shortly) preserved by the same unitary symmetries, but with both the states and symmetry generators written in terms of the generalised ladder operators $A_i$ and $A_i^{\dagger}$. 

\begin{equation} \notag
 \begin{split}
     S_1^u=&{\nu_{e}}\omega_1\omega_1^\dagger + \\  \overline{d^r}{A_1^\dagger}\omega_1\omega_1^\dagger + &\overline{d^g}{A_2^\dagger}\omega_1\omega_1^\dagger + \overline{d^b}{A_3^\dagger}\omega_1\omega_1^\dagger + \\ u^r{A_3^\dagger}{A_2^\dagger}\omega_1\omega_1^\dagger + &u^g{A_1^\dagger}{A_3^\dagger}\omega_1\omega_1^\dagger + u^b{A_2^\dagger}{A_1^\dagger}\omega_1\omega_1^\dagger + \\ &e^+{A_3^\dagger}{A_2^\dagger}{A_1^\dagger}\omega_1\omega_1^\dagger,
\end{split}
\qquad 
\begin{split}
     S_1^d=&{\overline{\nu}_{e}}\omega_1^\dagger\omega_1 + \\ {d}^r{A_1}\omega_1^\dagger\omega_1 + &{d}^g{A_2}\omega_1^\dagger\omega_1 +  {d}^b{A_3}\omega_1^\dagger\omega_1 + \\ \overline{u^r}{A_3}{A_2}\omega_1^\dagger\omega_1 + &\overline{u^g}{A_1}{A_3}\omega_1^\dagger\omega_1 + \overline{u^b}{A_2}{A_1}\omega_1^\dagger\omega_1 + \\ &e^-{A_3}{A_2}{A_1}\omega_1^\dagger\omega_1,
\end{split}
\end{equation}
where $\omega_1:=A_1A_2A_3$.

The $SU(3)_C$ generators are now a direct generalization of those in Section \ref{Furey}, with $\alpha_i$ replaced by $A_i$:
\begin{align} \notag
     \Lambda_1 &= -A_2^\dagger A_1 - A_1^\dagger A_2 \notag \hspace{2em}
     \Lambda_2 = iA_2^\dagger A_1 -iA_1^\dagger A_2 \notag \hspace{2em}
     \Lambda_3 = A_2^\dagger A_2 - A_1^\dagger A \notag\\ 
     \Lambda_4 &= -A_1^\dagger A_3 - A_3^\dagger A_1 \notag \hspace{2em}
     \Lambda_5 = -iA_1^\dagger A_3 + iA_3^\dagger A_1 \notag \hspace{2em}
     \Lambda_6 = -A_3^\dagger A_2 - A_2^\dagger A_3 \notag\\
     \Lambda_7 &= iA_3^\dagger A_2 - iA_2^\dagger A_3 \notag \hspace{1em}
     \Lambda_8 = -\frac{1}{\sqrt{3}}(A_1^\dagger A_1 + A_2^\dagger A_2 - 2A_3^\dagger A_3), \notag
 \end{align}
Furthermore, the number generator can again be used to define a $U(1)$ generator that assigns the correct electric charges to all the states:
\begin{eqnarray}\label{chargeQ}
    Q=\frac{N}{3}=\frac{1}{3}(A_1^{\dagger}A_1+A_2^{\dagger}A_2+A_3^{\dagger}A_3).
\end{eqnarray}
 
\subsection{Generating two additional generations from $S_3$ automorphisms}

Having constructed one generation of electrocolor states, we now proceed to apply $\psi$ to the ladder operators eqns. (\ref{newladder1})-(\ref{newladder3}), and subsequently the basis states of the minimal left ideals, in order to generate two additional generations. 


The first thing to check is that the $S_3$ automorphisms of $\bb{S}$ carry over the automorphisms of the left multiplication algebra $\bb{C}\ell(8)$. Given the generating basis $e_0,...,e_8$ for $\bb{C}\ell(8)$, it is readily verified that
\begin{eqnarray}
    \psi(e_i)\psi(e_i)=\psi(e_i^2)=-1,\\
    \psi(e_i)\psi(e_j)+\psi(e_i)\psi(e_j)=0.
\end{eqnarray}
Subsequently, $\psi$ extends to an automorphism of the entire $\bb{C}\ell(8)$ algebra (see Lemma 9.7 of \cite{harvey1990spinors}). A similar argument holds for the order two $S_3$ automorphism $\epsilon$ of $\bb{S}$, which likewise extends to an automorphism of $\bb{C}\ell(8)$.
It can furthermore be checked that:
\begin{eqnarray}
    \psi(e_{i+8})\psi(e_{i+8})=\psi(e_{i+8}^2)=-1,\\
    \psi(e_{i+8})\psi(e_{j+8})=-\psi(e_{i+8})\psi(e_{j+8}).
\end{eqnarray}
Finally, one also find that
\begin{eqnarray}
    \psi^2(e_i)+\psi(e_i)+e_i=0,\\
    \psi^2(e_{i+8})+\psi(e_{i+8})+e_{i+8}=0,
\end{eqnarray}
indicating that $\psi^2(e_i),\; \psi(e_i)$ and $e_i$ (as well as $\psi^2(e_{i+8}),\; \psi(e_{i+8})$ and $e_{i+8}$) are not linearly independent. These conditions however are not satisfied by more general $\bb{C}\ell(8)$ multivectors (nor for $e_8$, which is fixed by $\psi$). 



Since the action of $\psi$ on $e_1,...,e_{15}$ as elements of $\bb{C}\ell(8)$ is known, we can via linearity establish the action of $\psi$ on the ladder operators. This gives the three sets of ladder operators: 
\begin{eqnarray}
    A_i^{\dagger}&=&\frac{1}{2\sqrt{2}}(e_i+ie_{i+4}+e_{i+8}+ie_{i+12}),\\
    A_i&=&\frac{1}{2\sqrt{2}}(-e_i+ie_{i+4}-e_{i+8}+ie_{i+12}),\\
    \psi(A_i^{\dagger})=B_i^{\dagger}&=&\frac{1}{2\sqrt{2}}(ae_i+iae_{i+4}+be_{i+8}+ibe_{i+12}),\\
    \psi(A_i)=B_i&=&\frac{1}{2\sqrt{2}}(-ae_i+iae_{i+4}-be_{i+8}+ibe_{i+12}),\\
    \psi^2(A_i^{\dagger})=C_i^{\dagger}&=&\frac{1}{2\sqrt{2}}(be_i+ibe_{i+4}+ae_{i+8}+iae_{i+12}),\\
    \psi^2(A_i)=C_i&=&\frac{1}{2\sqrt{2}}(-be_i+ibe_{i+4}-ae_{i+8}+iae_{i+12}),
\end{eqnarray}
where $i=1,2,3$, and
\begin{eqnarray}
    a=\frac{\sqrt{3}-1}{2},\qquad b=\frac{-\sqrt{3}-1}{2},
\end{eqnarray}
satisfying $a^2+b^2=2,\; a+b=-1,\; ab=-1/2$.

The three sets of ladder operators are not linearly independent since
\begin{eqnarray}\label{conditionladder}
    \psi^2(A_i^{\dagger})+\psi(A_i^{\dagger})+A_i^{\dagger}=0.
\end{eqnarray}

The two additional sets of ladder operators $\{B_i^{\dagger},B_i\}$ and $\{C^{\dagger},C_i\}$ generated in this manner likewise constitute Witt bases for $\bb{C}\ell(6)$, satisfying the same anticommutation relations as $\{A_i^{\dagger},A_i\}$, and we proceed to construct an additional pair of minimal left ideals of $\bb{C}\ell(6)$ for each of the two additional sets of ladder operators $\{B_iB_i^{\dagger}\}$ and $\{C_i,C_i^{\dagger}\}$. We interpret these additional pairs of minimal ideals as representing the second and third generation of electrocolor states:
\begin{equation} \notag
 \begin{split}
     S_2^u=&{\nu_{\mu}}\omega_2\omega_2^\dagger + \\  \overline{c^r}{B_1^\dagger}\omega_2\omega_2^\dagger + &\overline{c^g}{B_2^\dagger}\omega_2\omega_2^\dagger + \overline{c^b}{B_3^\dagger}\omega_2\omega_2^\dagger + \\ s^r{B_3^\dagger}{B_2^\dagger}\omega_2\omega_2^\dagger + &s^g{B_1^\dagger}{B_3^\dagger}\omega_2\omega_2^\dagger + s^b{B_2^\dagger}{B_1^\dagger}\omega_2\omega_2^\dagger + \\ &\mu^+{B_3^\dagger}{B_2^\dagger}{B_1^\dagger}\omega_2\omega_2^\dagger
\end{split}
\qquad 
\begin{split}
     S_2^d=&{\overline{\nu}_{\mu}}\omega_2^\dagger\omega_2 + \\ {c}^r{B_1}\omega_2^\dagger\omega_2 + &{c}^g{B_2}\omega_2^\dagger\omega_2 +  {c}^b{B_3}\omega_2^\dagger\omega_2 + \\ \overline{s^r}{B_3}{B_2}\omega_2^\dagger\omega_2 + &\overline{s^g}{B_1}{B_3}\omega_2^\dagger\omega_2 + \overline{s^b}{B_2}{B_1}\omega_2^\dagger\omega_2 + \\ &\mu^-{B_3}{B_2}{B_1}\omega_2^\dagger\omega_2
\end{split}
\end{equation}

\begin{equation} \notag
 \begin{split}
     S_3^u=&{\nu_{\tau}}\omega_3\omega_3^\dagger + \\  \overline{b^r}{C_1^\dagger}\omega_3\omega_3^\dagger + &\overline{b^g}{C_2^\dagger}\omega_3\omega_3^\dagger + \overline{b^b}{C_3^\dagger}\omega_3\omega_3^\dagger + \\ t^r{C_3^\dagger}{C_2^\dagger}\omega_3\omega_3^\dagger + &t^g{C_1^\dagger}{C_3^\dagger}\omega_3\omega_3^\dagger + t^b{C_2^\dagger}{C_1^\dagger}\omega_3\omega_3^\dagger + \\ &\tau^+{C_3^\dagger}{C_2^\dagger}{C_1^\dagger}\omega_3\omega_3^\dagger
\end{split}
\qquad 
\begin{split}
     S_3^d=&{\overline{\nu}_{\tau}}\omega_3^\dagger\omega_3 + \\ {b}^r{C_1}\omega_3^\dagger\omega_3 + &{b}^g{C_2}\omega_3^\dagger\omega_3 +  {b}^b{C_3}\omega_3^\dagger\omega_3 + \\ \overline{t^r}{C_3}{C_2}\omega_3^\dagger\omega_3 + &\overline{t^g}{C_1}{C_3}\omega_3^\dagger\omega_3 + \overline{t^b}{C_2}{C_1}\omega_3^\dagger\omega_3 + \\ &\tau^-{C_3}{C_2}{C_1}\omega_3^\dagger\omega_3
\end{split}
\end{equation}
Here, $\omega_2:=B_1B_2B_3,\;\omega_3:=C_1C_2C_3$.

That is the order three automorphism $\psi$ of the finite group $S_3$ can be used to construct exactly two additional generations of color states. This automorphism then permutes between the three generations. 

On the other hand, the order two automorphism $\epsilon$ does not generate transformations between the three generations. Applying $\epsilon$ to the ladder operators $A_i$ and $A_i^{\dagger}$ generates an complementary set of ladder operators satisfying the same anticommutation relations as $A_i$ and $A_i^{\dagger}$ (since $\epsilon$ is an automorphism of $\bb{C}\ell(8)$), and hence can likewise be used to construct minimal ideals. The automorphism $\epsilon$ can then be used to incorporate an additional degree of freedom (perhaps chirality, handedness, or spin). It is not immediately clear at present however what the most natural physical interpretation for $\epsilon$ is.


\subsection{$SU(3)_C$ color symmetries of three generations}

The anticommutators between ladder operators belonging to different generations are not the standard fermionic anticommutation relations. That is
\begin{eqnarray}
    \{A_i,B_j\}w\neq 0,\qquad \{B_i,C_j\}w\neq 0,\qquad \{C_i,A_j\}w\neq 0, \quad\textrm{whenever}\;i\neq j
\end{eqnarray}
and likewise for the other anticommutation relations. Surprisingly however, it turns out that all three generations of states transform correctly under a single copy of $SU(3)$, which may be generated from any of the three sets of $\bb{C}\ell(6)$ ladder operators. We will here use the $SU(3)_C$ generators from Subsection \ref{onegenS}. One then finds for example that
\begin{eqnarray}
    \left[\Lambda_1, {A_1}\omega_1^\dagger\omega_1 \right]= {A_2}\omega_1^\dagger\omega_1,\quad
    \left[\Lambda_1, {B_1}\omega_1^\dagger\omega_1 \right]= {B_2}\omega_1^\dagger\omega_1,\quad
    \left[\Lambda_1, {C_1}\omega_1^\dagger\omega_1 \right]= {C_2}\omega_1^\dagger\omega_1.
\end{eqnarray}
A single $SU(3)$ generator thus correctly transforms the equivalent states of each generation
\begin{eqnarray}
    \Lambda_1: d^r\rightarrow d^g,\quad c^r\rightarrow c^g,\quad t^r\rightarrow t^g.
\end{eqnarray}
The same holds true for the other $SU(3)$ generators and states. Despite having three generations of fermions, only one copy of $SU(3)$, and hence one generation of gauge bosons is needed in the present construction. 

On the other hand, one finds that the number operator only assigns the correct electric charge for one of the generations. This same issue was encountered in \cite{furey2014generations}. Thus although the number operator can be used to obtain the correct electric charge within the context of a single generation model, this approach fails to work in a three generation model.  A generalized action to obtain the correct electric charges is presented in \cite{furey2018three}. This approach, which makes extensive use of both left and right projectors would work for the present model as well. However, it remains unclear why this action takes the particular form that it does, and so this issue will be revisited at a later time once the weak (or electroweak) symmetry has been included within the present sedenion model.

\subsection{Linear dependence of minimal ideal states}
Although we have been able to construct three pairs of minimal ideals to represent three generations of fermions, these minimal ideals are not linearly independent. Indeed,  one might expect a certain degree of overlap of the minimal ideals, given the condition eqn. (\ref{conditionladder}) above.

We propose (as in \cite{gillard2019three} and  \cite{manogue2022octions}) that the inevitable overlap of the three generations could form a basis for including quark mixing and neutrino oscillations into the present algebraic model based on sedenions. However, a detailed investigation into the feasibility of this proposal is beyond the scope of this article, and requires first an understanding of the weak interaction within the present algebraic context.

Nonetheless, some initial calculations (carried out using Mathematica) indicate that the (anti) down-type quarks across generations are linearly dependent. That is:
\begin{eqnarray}
    A_i\omega_1^{\dagger}\omega_1+B_i\omega_2^{\dagger}\omega_2+C_i\omega_3^{\dagger}\omega_3&=&0,\\
    A_i^{\dagger}\omega_1\omega_1^{\dagger}+B_i^{\dagger}\omega_2\omega_2^{\dagger}+C_i^{\dagger}\omega_3\omega_3^{\dagger}&=&0,\quad i=1,2,3
\end{eqnarray}
On the other hand, all 18 of the (anti) up-type quarks turn out to be linearly independent. This might be related to the fact that the Yukawa couplings, and subsequent mass matrix for either the up-type or down-type quarks can be taken to be diagonal, but not both. Subsequently, one can say without loss of generality that only the down-type quarks mix via the CKM matrix

\section{Outlook}

The model presented here is clearly far from complete as our focus has been restricted to the $SU(3)_C$ color gauge symmetry, and we have not considered the remaining internal symmetries nor the spacetime symmetries. 

One way to include chiral $SU(2)_L$ states it is via a copy of $\bb{C}\ell(4)$ \cite{furey2018demonstration}. In \cite{furey2016standard} it is shown that the right actions of $\bb{C}\otimes\bb{H}$ together with the right actions of the $\bb{C}\ell(6)$ nilpotents $\omega$ and $\omega^{\dagger}$ generate the required $\bb{C}\ell(4)$ to represent chiral $SU(2)_L$ states in terms of two four-dimensional minimal ideals. The resulting model is then based on $\bb{C}\ell(10)$ \cite{furey20183,gresnigt2020standard,todorov2023octonion}. In our present construction, we have restricted ourselves to a $\bb{C}\ell(6)$ subalgebra of the full $\bb{C}\ell(8)$ left multiplication algebra. The thus far unused $\bb{C}\ell(2)$ could be combined with the right actions of the nilpotents $\omega_i$ and $\omega^{\dagger}_i$ to generate $\bb{C}\ell(4)$ without the need to invoke $\bb{H}$. This approach is currently being investigated by the authors.

It is well known that $\bb{C}\otimes\bb{H} \cong \CCl(2) \cong SL(2,\bb{C})$. By complementing the present model sedenion model with a factor of $\bb{H}$, it should be possible to include the Lorentz (spacetime) symmetries into the model. Specifically, it was shown in \cite{furey2016standard} that Weyl-, Dirac- and Majorana-spinors can all be represented in terms of the ideals of $\bb{C} \otimes \bb{H}\cong \CCl(2)$. The spacetime symmetries would then be identical for all three generations, a desirable feature.

By choosing a complex structure within $\bb{O}$, the ten dimensional spacetime described by $SL(2,\bb{O})$ is reduced to four dimensional spacetime $SL(2,\bb{C})$ \cite{manogue1999dimensional,dray2015geometry}. It would be interesting to see whether $SL(2,\bb{O})$ spacetime can be extended to an 18 dimensional spacetime $SL(2,\bb{S})$ which is then broken to $SL(2,\bb{H})$ via our quaternionic structure inside $\bb{S}$, and subsequently to $SL(2,\bb{C})$ by choosing a complex structure inside $\bb{H}$. 

The construction of three generations presented here bears some intriguing resemblance to the three generation construction in \cite{furey2014generations,furey2018three}. In our case, all three generations reside in a $\bb{C}\ell(6)$ subalgebra of $\bb{C}\ell(6)$. Likewise in \cite{furey2014generations}, the 48 degrees of freedom that remain in $\bb{C}\ell(6)$ once two representations of the Lie algebra $su(3)$ have been accounted for are shown under the action of $SU(3)$ to transform as three generations of leptons and quarks. The number operator in this construction no longer assigns the correct electric charges to the states however, an issue that was overcome in a later paper \cite{furey2018three}. How to include $U(1)_{em}$ into the sedenion model, as well as a detailed understanding of how these two complementary constructions are related remains to be worked out.

Likewise, it remains to be investigated in detail how the sedenion model proposed here relates to the constructions of three generations based on the exceptional Jordan $J_3(\bb{O})$, in which each of the three octonions in $J_3(\bb{O})$ is associated with one generation via the three canonical $J_2(\bb{O})$ subalgebras of $J_3(\bb{O})$ \cite{dubois2016exceptional,dubois2019exceptional,todorov2018octonions,todorov2018deducing,boyle2020standard,boyle2020standard2}.


Given the richer algebraic structure provided by $\bb{S}$, it may be possible to describe additional features of the SM. The $S_3$ automorphism of order three not only generates two additional generations, but also facilitate transformations between physical states of different generations. This suggests that $S_3$ could perhaps be used as a basis for including quark mixing and neutrino oscillations. This is something that has not yet been considered within the context of division algebras. It is interesting that several authors have proposed $S_3$ extensions of the SM to explain the hierarchy of quark masses and mixing in the SM \cite{kubo2003flavor,kubo2004higgs,kubo2004majorana,kubo2005minimal,mondragon2007lepton}.


\section{Discussion}
We have argued that the CD algebra $\bb{S}$ provides a suitable algebraic structure to describe three generations. Our focus has been restricted to the $SU(3)_C$ color gauge symmetry. Three intersecting $\bb{O}$ subalgebras of $\bb{S}$ are used to construct a Witt basis for $\bb{C}\ell(6)$. Two minimal left ideals are then use to represent one generation of electrocolor states. Subsequently, the $S_3$ automorphism of order three of $\bb{S}$ is then applied to the $\bb{C}\ell(6)$ Witt basis in order to obtain exactly two additional generations of color states, but not electrocolor states.  

In \cite{dixon2013division,furey2016standard}, $\bb{C}\ell(6)$ arises as the left multiplication algebra of $\bb{C}\otimes\bb{O}$. Instead, in our approach $\bb{C}\ell(6)$ does not arise as the multiplication algebra of a single octonion algebras, but rather from three intersection octonion subalgebras of the sedenions, with each octonion subalgebra contributing a $\bb{C}\ell(2)$ factor. The model presented here overcomes two limitations of a previous three generation model based on sedenions. First, only a single copy of $SU(3)_C$ correctly transforms the states of all three generations. We thereby avoid introducing three generations of gauge bosons. Additionally in the present model, the order three $S_3$ automorphisms of $\bb{S}$ if given a clear physical interpretation as a family symmetry, responsible for generating two additional generations from the first.  

One compelling reason to consider division algebras as the foundational mathematical input from which to generate SM particle multiplets and gauge symmetries, is that there are only four of them. Since the CD process generates an infinite series of algebras, one might question whether going beyond the division algebras and including $\bb{S}$ is a wise idea, or if it opens the door to considering ever larger algebras. The derivation algebra for all CD algebras $\bb{A}_n$, $n\geq 3$ is equal to $\mathfrak{g}_2$ however \cite{schafer1954algebras}. Furthermore, at least for the cases $n=4,5,6$, the automorphism group of each successive CD algebra only picks up additional factors of $S_3$ \cite{brown1967generalized}. It therefore seems unlikely that CD algebras beyond $\bb{A}_4=\bb{S}$ will provide additional physical insight. As an interesting aside, the sphere $S^{15}$, associated with the imaginary (pure) sedenions, is the largest sphere to appear in any of the four Hopf fibrations.

\subsection*{Acknowledgements}
The authors wish to thank Alessio Marrani insightful discussions and detailed feedback on earlier drafts of this work. 

\section*{Appendix A: Sedenion multiplication table}

Here we provide the multiplication table for the Cayley-Dickson sedenions that we use, written in terms of the indices ($i\equiv e_i)$:

\begin{center}
\begin{table}[h!]
\resizebox{\columnwidth}{!}{%
\begin{tabular}{|c|| c c c c c c c c c c c c c c c c|} 
 \hline
 & 0 & 1 & 2 & 3 & 4 & 5 & 6 & 7 & 8 & 9 & 10 & 11 & 12 & 13 & 14 & 15\\ [0.5ex] 
 \hline\hline
 0 & 0 & 1 & 2 & 3 & 4 & 5 & 6 & 7 & 8 & 9 & 10 & 11 & 12 & 13 & 14 & 15\\ 
 
 1 & 1 & -0 & 3 & -2 & 5 & -4 & -7 & 6 & 9 & -8 & -11 & 10 & -13 & 12 & 15 & -14\\

 2 & 2 & -3 & -0 & 1 & 6 & 7 & -4 & -5 & 10 & 11 & -8 & -9 & -14 & -15 & 12 & 13\\
 
 3 & 3 & 2 & -1 & -0 & 7 & -6 & 5 & -4 & 11 & -10 & 9 & -8 & -15 & 14 & -13 & 12\\

 4 & 4 & -5 & -6 & -7 & -0 & 1 & 2 & 3 & 12 & 13 & 14 & 15 & -8 & -9 & -10 & -11\\ [1ex] 

 5 & 5 & 4 & -7 & 6 & -1 & -0 & -3 & 2 & 13 & -12 & 15 & -14 & 9 & -8 & 11 & -10\\ [1ex] 
 
 6 & 6 & 7 & 4 & -5 & -2 & 3 & -0 & -1 & 14 & -15 & -12 & 13 & 10 & -11 & -8 & 9\\ [1ex] 

 7 & 7 & -6 & 5 & 4 & -3 & -2 & 1 & -0 & 15 & 14 & -13 & -12 & 11 & 10 & -9 & -8\\ [1ex] 

 8 & 8 & -9 & -10 & -11 & -12 & -13 & -14 & -15 & -0 & 1 & 2 & 3 & 4 & 5 & 6 & 7\\ [0.5ex] 
 
 9 & 9 & 8 & -11 & 10 & -13 & 12 & 15 & -14 & -1 & -0 & -3 & 2 & -5 & 4 & 7 & -6\\ 
 
 10 & 10 & 11 & 8 & -9 & -14 & -15 & 12 & 13 & -2 & 3 & -0 & -1 & -6 & -7 & 4 & 5\\
 
 11 & 11 & -10 & 9 & 8 & -15 & 14 & -13 & 12 & -3 & -2 & 1 & -0 & -7 & 6 & -5 & 4\\
 
 12 & 12 & 13 & 14 & 15 & 8 & -9 & -10 & -11 & -4 & 5 & 6 & 7 & -0 & -1 & -2 & -3\\
 
 13 & 13  & -12 & 15 & -14 & 9 & 8 & 11 & -10 & -5 & -4 & 7 & -6 & 1 & -0 & 3 & -2\\ [1ex] 
 
 14 & 14 & -15 & -12 & 13 & 10 & -11 & 8 & 9 & -6 & -7 & -4 & 5 & 2 & -3 & -0 & 1\\ [1ex] 
 
 15 & 15 & 14 & -13 & -12 & 11 & 10 & -9 & 8 & -7 & 6 & -5 & -4 & 3 & 2 & -1 & -0\\ [1ex] 
 \hline

\end{tabular}
}
\caption{Sedenion multiplication table, generated from $e_1, e_2, e_4$ and $e_8$ corresponding to the new $\bb{C}$, $\bb{H}$, $\bb{O}$ and $\bb{S}$ imaginary units respectively.}
\end{table}
\end{center}


\bibliography{Bibliography2}  
\bibliographystyle{unsrt}  

\end{document}